\documentclass[a4paper,11pt]{article}
\pdfoutput=1 

\usepackage{jheppub} 

\usepackage{amsmath}
\usepackage{amssymb}
\usepackage{appendix}
\usepackage{multirow} 
\usepackage{rotating}
\usepackage{diagbox}
\usepackage{slashed}
\usepackage{float}
\usepackage[dvipsnames]{xcolor}
\usepackage{cancel}
\usepackage{tabularx}
\usepackage{bbold}
\usepackage[capitalize]{cleveref}
\usepackage[normalem]{ulem}
\usepackage{soul}
\usepackage{mathtools}

\newcommand{\bra}[1]{\mathopen{}\left<#1\right|}
\newcommand{\ket}[1]{\mathopen{}\left|#1\right>}
\newcommand{\ang}[1]{\mathopen{}\left<#1 \right>}
\newcommand{\braket}[2]{\mathopen{}\left<#1 \middle|#2\right>}
\newcommand{\amp}{\mathcal{A}}
\renewcommand{\doteq}{\overset{\cdot}{=}}
\renewcommand{\refeq}[1]{(\ref{#1})}
\newcommand{\beq}{\begin{equation}}
\newcommand{\eeq}{\end{equation}}
\newcommand{\nn}{\nonumber \\}
\newcommand{\bal}{\begin{align}}
\newcommand{\eal}{\end{align}}

\title{Gauge theories from scattering amplitudes with minimal assumptions}

\author[a]{Renato M. Fonseca,}
\author[a]{Clara Hernández-García,}
\author[b]{Javier M. Lizana,}
\author[a]{Manuel~Pérez-Victoria}

\affiliation[a]{Departamento de Física Teórica y del Cosmos, Universidad de Granada, 18071 Granada, Spain}

\affiliation[b]{Instituto de F\'isica Te\'orica UAM/CSIC, Nicolas Cabrera 13-15, 28049 Madrid, Spain}

\affiliation[b]{Departamento F\'isica Aplicada a las Ingenier\'ias Aeron\'autica y Naval, E.T.S.I. Aeron\'autica y del Espacio, Universidad Polit\'ecnica de Madrid, 28040 Madrid, Spain}

\emailAdd{renatofonseca@ugr.es}
\emailAdd{clarahdez@ugr.es}
\emailAdd{javier.mlizana@upm.es}
\emailAdd{mpv@ugr.es}

\abstract{We revisit the emergence of a Yang-Mills symmetry in theories with massless spin 1 particles from fundamental physical properties of scattering amplitudes. In the standard proofs, some symmetry and reality properties of the coupling constants in three-point amplitudes are assumed. These properties cannot be justified using only three-point amplitudes but we show that they arise as consequences of the consistent factorization  of four-particle amplitudes, for particular choices of the particle basis. This applies to self-interactions of massless spin 1 particles and also to their interactions with spin 0 and 1/2 particles. CP invariance is a derived property, not an additional assumption. The situation for gravity interactions is analogous and it is dealt with in the same fashion.}

\makeatletter
\gdef\@fpheader{}
\makeatother

\begin{document} 
\maketitle
\flushbottom

\section{Introduction  }
\label{sec:Intro}

The modern on-shell approach to scattering amplitudes has shown that
several classes of interacting quantum field theories can be reconstructed
by applying a minimal set of physical principles directly to the
$S$ matrix. In particular, unitarity, locality and Poincaré invariance
severely constrain --- and in some cases completely fix --- the
form of allowed interactions. The case of massless spin-1 and spin-2
particles (gluons and gravitons) is particularly noteworthy: it has
been advocated that these simple on-shell principles lead unavoidably
to the Yang-Mills construction, and, in the
case of gravity, to universal couplings to all particles~\cite{Benincasa:2007xk,Schuster:2008nh,McGady:2013sga}.

The usual story for the spin 1 case can be summarized as follows.
The kinematic dependence of three-point amplitudes of massless particles
is completely determined by their helicities, and a single gluon is
readily shown to have no self-interactions. More generally, in a theory
with several spin 1 particles distinguished by color indices, one
can denote by a set of constants $f_{ab}^{c}$ and $\bar{f}^{ab}_c$ the coupling strength
of in-going gluons with colors $abc$ and helicities $++-$ and $--+$, respectively. In the
standard presentation the $f$ and $-\bar{f}$ tensors are taken to be equal, imaginary and totally antisymmetric.\footnote{The sign in $-\bar f$ appears due to our conventions, to be specified later. Often the three-point amplitude is written with an $ig$ prefactor,
$\mathcal{A}\left(1_{a}^{+},2_{b}^{+},3_{c}^{-}\right)=igf_{ab}^{c}\cdots$,
in which case $f$ is said to be real ($g$ being a real coupling
constant).} Then, by requiring correct factorization
of the four-gluon amplitudes one concludes \cite{Benincasa:2007xk,Schuster:2008nh,McGady:2013sga} that $f$
must obey the constraint\footnote{In this work, we indicate sums of repeated indices explicitly.}
\begin{equation}
\sum_{x}\left(f_{bc}^{x}f_{ax}^{d}+f_{ca}^{x}f_{bx}^{d}+f_{ab}^{x}f_{cx}^{d}\right)=0\,,\label{eq:JI}
\end{equation}
which is the Jacobi identity for the structure constants $f_{ab}^{c}$
of a Lie group.

The equality $\bar{f}^{ab}_c=-f_{ab}^c$ can be justified by assuming invariance under parity P, as done in ~\cite{Benincasa:2007xk,Schuster:2008nh,McGady:2013sga}. The reality of $-if_{ab}^{c}$ also follows from this assumption, since, as long as CPT symmetry is present, it implies invariance under time reversal T.\footnote{For massless particles, the distinction between P and CP is essentially linguistic. Most of the time we will speak of CP.}
Regarding the symmetry
of $f$ under index permutations, the relevant kinematic factor is
antisymmetric under exchange of the momenta of the $+$ helicity gluons;
Bose symmetry then forces $f_{ab}^{c}$ to be antisymmetric under
the exchange $a\leftrightarrow b$. However,
the standard understanding (see for instance~\cite{Benincasa:2007xk,Schuster:2008nh,McGady:2013sga,Schwartz:2013pla,Arkani-Hamed:2017jhn,Cheung:2017pzi}) is that $f$ is antisymmetric
under exchange of all indices, including $c$.

We argue, however, that it is not possible to make any general connection between
what are fundamentally two different sets of indices: the colors of
in-going particles with positive helicity ($ab$), and the color of
an in-going particle with negative helicity ($c$). Note that crossing
symmetry~\cite{Eden:1966dnq,Hannesdottir:2022bmo} relates in- and out-states at the amplitude level, not in-
and in-states. So, even if it helps in reducing the different types of indices from four (in $+$, in $-$, out $+$, out $-$) to two (in $+$, in $-$), it does not reduce the number to one. Besides, the full antisymmetry of $f$ cannot be a
basis-independent property, as a general unitary rotation $U$ in color
space,
\begin{equation}
f_{ab}^{c}\rightarrow\left(f^{\prime}\right)_{ab}^{c}:=\sum_{a^{\prime}b^{\prime}c^{\prime}}U_{a}^{\;a^{\prime}}U_{b}^{\;b^{\prime}}\left(U_{c}^{\;c^{\prime}}\right)^{*}f_{a^{\prime}b^{\prime}}^{c^{\prime}}\,,\label{eq:U-rotation}
\end{equation}
 spoils this feature if it ever was present in some basis. Such transformations of one-particle states, which are allowed in a quantum theory and are described below more precisely, do not change the kinematics due to the degeneracy of masses.

Concerning parity or time reversal, it would be interesting to study theories
with spin 1 particles in full generality, and for that one must set
aside such assumptions. Note also that even if parity implies $\bar{f}=-f$ and the reality of $-if$ in a basis associated to its precise definition, this does not help in proving the complete antisymmetry of $f$ in that basis (or in another one).

Rejecting the assumptions of complete antisymmetry and of parity or time-reversal symmetry, what one does get from the correct factorization of four-gluon amplitudes is the condition 
\begin{equation}
\sum_{x}\left(\bar{f}^{xa}_{d} f_{bx}^{c}+\bar{f}^{ac}_{x} f_{bd}^{x}+\bar{f}^{ax}_{b} f_{dx}^{c}\right)=0\label{eq:MJI-0} \, .
\end{equation}
Assuming only CPT symmetry, in a sense that will be specified below, this can be written as
\begin{equation}
\sum_{x}\left[\left(f_{xa}^{d}\right)^{*}f_{bx}^{c}+\left(f_{ac}^{x}\right)^{*}f_{bd}^{x}+\left(f_{ax}^{b}\right)^{*}f_{dx}^{c}\right]=0 \, , \label{eq:MJI-1}
\end{equation}
which, due to its similarity to equation (\ref{eq:JI}), we shall
call the modified Jacobi identity. On one hand, it is far from obvious
(and thus the extra assumptions in the literature) that this equation
enforces a Lie algebra structure on the $f$ tensor. On the other
hand, even if it did, there is no obvious reason why there should
be any basis where $f$ is fully antisymmetric. On this last point,
we note that there are perfectly valid Lie algebras for which $f$
does not have this property, the simplest of which is the two-dimensional
affine algebra spanned by two elements $A$ and $B$ and with a commutator
$\left[A,B\right]=A$.

At this stage, let us make a small detour from the on-shell
approach, and recall the properties of the $f$ tensor in the Lagrangian
formalism~\cite{Weinberg:1996kr}. In a gauge theory, the massless spin 1 particles are created by vector bosons, which are real fields sitting
in the adjoint representation of the gauge group, parametrized by real parameters $\alpha^a$, and which is determined
by the imaginary structure constants $f$:
\begin{equation}
\delta_\alpha \left(A_{\mu}^{a}\right)= i\sum_{b,c}\alpha^{b}f_{bc}^{a}A_{\mu}^{c} .
\end{equation} 
The reality of the vector bosons already implies that the associated particles of both helicities couple with the same coupling strength. Furthermore, gauge invariance requires that these gauge transformation leave invariant the metric in color space that is used to construct the kinetic
term of the vector bosons. The fact that the metric in the kinetic term is both invariant and positive definite has in turn two important implications: first, there is a basis in which the $f$ tensor is completely antisymmetric; second, the Lie algebra of the gauge group is necessarily reductive,
i.e. a direct sum of a semisimple subalgebra and an abelian one.

The stakes are therefore high: Can one have a theory of massless spin-1
particles that does not admit any underlying gauge group at all, yet
still satisfies unitarity, locality and Poincaré invariance? Even
if the answer is no and a gauge group is unavoidable, must its Lie
algebra be reductive, or is there perhaps some overlooked loophole
in the Lagrangian formalism that makes it possible to have more general
symmetry groups? And finally, could gluon self-interactions violate CP symmetry?

One tantalizing possibility is that a rotation (\ref{eq:U-rotation})
in color space can be used to cast $f$ in a more desirable form.
However, a simple parameter-counting casts doubt on this: for example,
with $n$ gluons one has only $\sim n^{2}$ degrees of freedom in
a unitary rotation $U$ and $\sim n^{3}$ complex phases in $f$.
Naively then, one cannot expect to be able to tune all these phases
and make $-if$ a real (or fully antisymmetric) tensor by a change of color basis.

Nevertheless, in this work, assuming only Poincar\'e invariance and the usual analyticity properties of tree-level scattering amplitudes, we prove the following results:
\begin{itemize}
\item There are bases --- which we will call real bases --- where $\bar{f}=-f$ and the
$-if$ tensor is real and fully antisymmetric.\footnote{Such bases define a real structure on the color space of one-particle states, which is necessary to identify it with the real color space of the gauge fields in the Lagrangian formalism.}
\item The Jacobi equation (\ref{eq:JI}) holds in any basis.
\end{itemize}
Hence, we show that $f_{ab}^{c}$ are indeed the structure constants
of a reductive Lie algebra; the field formalism of gauge theories has no loophole. 

Finally, the concern expressed above with the reality and symmetry
of the self-couplings of gluons also extends to other couplings: (1)
interactions of gluons with scalars and fermions, (2) self-interactions
of gravitons, (3) interactions of gravitons with particles with lower
spins. In this work we revisit them all and conclude that it is always
possible to find a basis where the standard assumptions for these couplings
hold. Then, the standard results follow: the gluon couplings furnish representations of the Lie algebra, there is a basis in which different species of gravitons decouple and the gravitons of both helicities couple universally to all particles. We also argue that properly-defined CP transformations leave the cubic interactions of gluons and gravitons invariant.

This work is structured as follows: Section~\ref{sec:gaugebosons} is devoted to  the self-interactions of massless spin 1 particles. In that section, we start in~\ref{sec:CPT} by carefully defining the relevant space of one-particle states, the basis transformations we will consider and the implications of CPT symmetry, which will be important throughout the paper. Then we define in~\ref{sec:cubic} the cubic couplings $f$ and examine which properties can be inferred from three-gluon amplitudes. After this, we derive in~\ref{sec:NAC} from the four-gluon amplitudes a crucial property of the couplings $f$, which we call the non-ambiguity condition. We briefly discuss the modified Jacobi identity in~\ref{sec:MJI}. To finish that section, we show in~~\ref{sec:hermiticity} that a basis exists in which the couplings $f$ are imaginary and totally antisymmetric, which then implies they are the structure constants of a reductive Lie algebra. In section~\ref{sec:matter} we generalize the analysis to the interactions of gluons with other massless particles: we show that there is a basis in which the couplings form hermitian matrices and conclude that they furnish representations of the Lie algebra identified in the previous section. In section~\ref{sec:symmetry} we describe explicitly how this algebraic structure translates into an essentially compact symmetry group. Section~\ref{sec:CP} is devoted to the emergence of CP symmetry in the gluon couplings. In section~\ref{sec:gravity}, we deal with the interactions of massless spin 2 particles in an analogous way. We conclude in section~\ref{sec:Conclusions}.

\section{Self-interactions of spin 1 particles}
\label{sec:gaugebosons}

\subsection{Color space and~CPT}
\label{sec:CPT}

We consider first a theory of one-particle states of massless particles of helicity $\pm 1$, which we will call gluons. Their internal degrees of freedom are represented by the finite-dimensional Hilbert spaces $\mathcal{H}^{+}$ and $\mathcal{H}^{-}$ associated with positive and negative helicity respectively. 
CPT is an antiunitary transformation that maps one-particle states of a given helicity into states of the opposite helicity,
\beq
\mathrm{CPT} \ket{\vec{p}_f^{\,h}} = \ket{\vec{p}_{\bar{f}}^{{\,-h}}}, \label{CPT1particle}
\eeq
where $f$ and $\bar{f}$ label specific internal degrees of freedom in $\mathcal{H}^\pm$, respectively, and we have taken into account that $|h|=1$ for gluons. 
CPT then defines an antilinear isometry between $\mathcal{H}^+$ and $\mathcal{H}^-$.
Given an orthonormal basis in $\mathcal{H}^+$, $\{|e_a\rangle \}$, we can use CPT to generate an orthonormal basis in $\mathcal{H}^-$, $\{|e^a\rangle\} = \{{\rm CPT}\,|e_a\rangle\}$. We will adopt such connected bases throughout this paper. Then, we can speak of particles of a given species (whose label will be called {\em color} for gluons and {\em flavor} in general)
with two possible helicities.
Using these bases, CPT acts on tensors by complex-conjugating their components, $[{\rm CPT}(T)]^{a_1\ldots a_n}_{b_1\ldots b_m}=(T_{a_1\ldots a_n}^{b_1\ldots b_m})^*$.
Notice that we use an upper index for the elements of the bases of negative helicity. This choice reflects the antilinearity of CPT, which implies that these elements transform differently under a change of basis, $|e_a\rangle\to  |e_b\rangle\, U^b_{a}$ while $|e^a\rangle\to  (U^{\dagger})^a_{b} \, |e^b\rangle$ where $U$ is an unitary matrix in order to preserve the orthonormality of the basis.
We see that bases of $\mathcal{H}^-$ transform like bases of one-forms or {\it bras} in the dual space $(\mathcal{H}^+)^*$.
Formally, we can identify $\mathcal{H}^-$ with the dual space $(\mathcal{H}^+)^*$ using CPT: the composition of CPT with the antilinear isometry between $\mathcal{H}^+$ and $(\mathcal{H}^+)^*$ induced by the inner product yields a linear isometry, $\mathcal{H}^- \sim (\mathcal{H}^+)^*$.

For multi-particle asymptotic states, which are necessary to define the S matrix, CPT maps {\it in} states into {\it out} states of opposite helicity, and vice-versa.
Then, CPT symmetry of the S matrix is the statement
\beq
\prescript{}{\mathrm{out}}{}\braket{\alpha}{\beta}_\mathrm{in} = \prescript{}{\mathrm{out}}{} \braket{\bar{\beta}}{\bar{\alpha}}_\mathrm{in},
\eeq
where $\alpha$ and $\beta$ label multi-particle states and the bar denotes the transformation of labels specified in~\refeq{CPT1particle} for each one-particle state. More generally, as customary in on-shell methods, we will consider scattering amplitudes for complex momenta and we will assume crossing symmetry~\cite{Eden:1966dnq,Hannesdottir:2022bmo}, which allows us to treat {\em out} states as {\em in} states of opposite helicity. 
Crossing all particles and analytically continuing to real momenta with positive energies corresponds to CPT invariance~\cite{Eden:1966dnq}. Assuming also hermitian analyticity~\cite{Eden:1966dnq}, crossing symmetry implies
\begin{align}
\prescript{}{\mathrm{out}}{}\braket{0}{\alpha}_\mathrm{in} & = \prescript{}{\mathrm{out}}{}\braket{\bar{\alpha}}{0}_\mathrm{in}   \nn
& = \prescript{}{\mathrm{out}}{} \braket{0}{\bar{\alpha}^*}_\mathrm{in}^* \, ,
\label{treeCPT}
\end{align}
where $\bar{\alpha}^*$ represents the multi-particle state obtained from $\alpha$ by flipping all the helicities and complex conjugating all the momenta.\footnote{Our results below only require~\refeq{treeCPT} to hold at the tree level. At the tree level, the property of Hermitian analyticity is just a generalization to complex momenta of the fact that the transition matrix is Hermitian at that order.} In the following we will refer to~\refeq{treeCPT} as the CPT relation.

\subsection{Three-gluon amplitudes}
\label{sec:cubic}

The renormalizable cubic interactions of three massless spin 1 particles with complex momenta are given by
\begin{align}
& \mathcal{A}\left(1_{a}^{+},2_{b}^{+},3_{c}^{-}\right) = f_{ab}^c \frac{[12]^3}{[13][23]}, \\
& \mathcal{A}\left(1_{a}^{-},2_{b}^{-},3_{c}^{+}\right) = -{\bar{f}}^{ab}_c \frac{\ang{12}^3}{\ang{13}\ang{23}}, 
\end{align}
for some color-dependent complex constants $f$ and $\bar{f}$. The CPT relation~\refeq{treeCPT} implies\footnote{We use a spinor phase convention such that $[p^*q^*]^* = \ang{qp}$ for arbitrary complex momenta $p$ and $q$.} ${\bar{f}}_c^{ab} = \left( f^{c}_{ab} \right)^*$. The notation of indices is as described above. Local, non-renormalizable interactions of three gluons with the same helicity are also possible. Such interactions do not interfere with arguments below, so for simplicity we assume they vanish in the following.

Bose symmetry and the antisymmetry under the exchange of the momenta of the helicity +1 gluons imply that the color tensor $f$ is antisymmetric under the exchange of the lower indices: $f_{ab}^c = - f_{ba}^c$. Similarly, and in agreement with the CPT relation, $\bar{f}_{ab}^c = - \bar{f}_{ba}^c$. But from these cubic interactions one cannot infer the antisymmetry of $f$ or $\bar{f}$ under the exchange of lower and upper indices, which would imply complete antisymmetry of these tensors. As a matter of fact, lower and upper indices correspond to particles with different helicity, i.e.\ to states living in different spaces, $\mathcal{H}_{\mathrm{in}}^+$ and $\mathcal{H}^-_{\mathrm{in}} \sim (\mathcal{H}^+_{\mathrm{in}})^*$. The linear version of the CPT isometry does not connect $\mathcal{H}^-_{\mathrm{in}}$ with $\mathcal{H}_{\mathrm{in}}^+$ but with its dual. Therefore, no such symmetry property is to be expected {\it a priori\/}.\footnote{This is in contrast with the non-renormalizable cubic couplings of three gluons with the same helicity, which are necessarily completely antisymmetric.} Furthermore, even if $f_{ab}^c$ were antisymmetric in one orthonormal basis, this property would not be preserved under a unitary change of basis, so it is at best a basis-depependent property. But no assumption about the color basis has been made besides orthonormality.

Note that CP invariance or, equivalently, T invariance, defined in their simplest form without color rotations or intrinsic phases, would imply that $f_{ab}^c$ is purely imaginary. This is again a basis-dependent statement. At any rate, at the level of cubic amplitudes CP does not give any new information about the antisymmetry of $f$. We will not assume CP symmetry in this paper, but we discuss the emergent CP invariance of gluon interactions in section~\ref{sec:CP}.

\subsection{The non-ambiguity condition}
\label{sec:NAC}

More properties about the couplings $f_{ab}^c$ can be derived from the requirement of a correct factorization of the four-point amplitude $\amp(1_a^+,2_b^+,3_c^-,4_d^-)$. We will first pay attention to the fact that, as stressed in~\cite{Schuster:2008nh}, in each $(ij)$ factorization channel, this should be true for the different configurations of complex momenta that are possible when $s_{ij} \to 0$, that is, both for $\ang{ij} \to 0$ and for $[ij] \to 0$. 

By little-group covariance, the amplitude must be of the form
\beq
\amp(1_a^+,2_b^+,3_c^-,4_d^-) = [12]^2\ang{34}^2 H_{abcd}(s_{12},s_{13}), \label{amplitudeform}
\eeq
for some function $H_{abcd}$.
Consider first the (12) channel. In the configuration with $\ang{12} \to 0$,  $[34] \to 0$,  $[12]\neq 0$, $\ang{34} \neq 0$, only one intermediate polarization contributes. In this limit, we must have
\begin{align}
\amp(1_a^+,2_b^+,3_c^-,4_d^-) & \sim \frac{1}{s_{12}} \sum_x  \amp(1_a^+,2_b^+,q_x^-) \amp(-q_x^+, 3_c^-,4_d^-) \nn
& = \sum_x f_{ab}^x(f_{cd}^x)^* \frac{[12]^2\ang{34}^2}{s_{12} s_{13}}, \label{residue++}
\end{align}
where $-q$ equals the sum of the momenta 1 and 2.
On the other hand, in the configuration with $[12] \to 0$, $\ang{12}\neq 0$ and/or $\ang{34} \to 0$, $[34]\neq 0$, we must have a vanishing residue, since we have chosen a holomorphic function for ($++-$) amplitudes and an antiholomorphic function for ($- -+$) amplitudes.\footnote{ Note that the configurations with $[12] \to 0$ and $[34] \to 0$, or $\ang{12} \to 0$ and $\ang{34} \to 0$, are only possible in the limit in which all the momenta are parallel (so all the Mandelstam variables vanish).} With different choices, the amplitude would be singular for $s_{12} \to 0$ in the limit of real momenta. So, the correct four point amplitude has the form~\refeq{residue++} in the limit $\ang{12} \to 0$, $[34] \to 0$ and vanishes when $[12]\to0$ or $\ang{34}\to0$. This is automatically consistent with the expression~\refeq{amplitudeform}, so we do not find any new constraint from this channel. Things get more interesting for the other two channels.

Consider next the (13) channel. Again, we distinguish different configurations of complex momenta compatible with $s_{13}\to0$. When  $\ang{13} \to 0$,  $[24] \to 0$,  $[13]\neq 0$, $\ang{24} \neq 0$, we must have
\begin{align}
\amp(1_a^+,2_b^+,3_c^-,4_d^-) & \sim \frac{1}{s_{13}} \sum_x \amp(1_a^+,3_c^-,q_x^+) \amp(-q_x^-,2_b^+,4_d^-) \nn
& = \sum_x  f_{xa}^c(f_{xd}^b)^* \frac{ [12]^2 \ang{34}^2 }{s_{13}s_{12}} \label{residue+-1} ,
\end{align}
while when $[13]\to 0$,  $\ang{24} \to 0$,  $\ang{13} \neq 0$, $[24] \neq 0$, we must have
\begin{align}
\amp(1_a^+,2_b^+,3_c^-,4_d^-) & \sim  \frac{1}{s_{13}} \sum_x  \amp(1_a^+,3_c^-,q_x^-) \amp(-q_x^+, 2_b^+,4_d^-) \nn
& = \sum_x (f_{xc}^a)^* f_{xb}^d \frac{ [12]^2 \ang{34}^2 }{s_{13}s_{12}} \label{residue+-2} .
\end{align}
Note that the dependence on kinematics is identical in both limits but the prefactor is not, in general. But the expression~\refeq{amplitudeform} shows that $\amp(1_a^+,2_b^+,3_c^-,4_d^-)$ has the same value in both limits. So, its consistent factorization requires $\sum_x f_{xa}^c(f_{xd}^b)^* = \sum_x (f_{xc}^a)^* f_{xb}^d$. 

The situation for the (14) channel is the same and consistent factorization demands $\sum_x f_{xa}^d (f_{xc}^b)^* = \sum_x (f_{xd}^a)^* f_{xb}^c$. Let us define the matrices $F_a$ with entries
\beq
(F_a)_{~b}^{c} = f_{ab}^c.
\eeq
Then, we have shown that consistent factorization requires the non-ambiguity condition (NAC)
\beq
\sum_a F_a \otimes F_a^\dagger = \sum_a F_a^\dagger \otimes F_a. \label{NAC}
\eeq

The same condition can also be obtained using constructiveness, i.e. recursion relations derived from a deformation with a complex variable and Cauchy's theorem. Let us show it explicitly using BCFW shifts~\cite{Britto:2004ap,Britto:2005fq,Arkani-Hamed:2008bsc}. We assume that there is no non-renormalizable interaction.

To calculate $\amp(1_a^+,2_b^+,3_c^-,4_d^-)$ from the cubic amplitudes we can choose to shift different pairs of legs, as long as we make sure that the residue at infinity does not contribute. The  shifts performed in the following fulfill this property. Let us first perform the calculation making the (1-4) shift
\beq
\bra{\tilde{1}}= \bra{1} +z \bra{4},~~  |\tilde{4}] = |4] - z |1], ~~z \in \mathbb{C},
\eeq
and keeping all the other momentum spinors unchanged. The $(\tilde{1}\tilde{4})$ channel does not contribute. The $(\tilde{1}2)$ channel gives a contribution
\beq
- \sum_x f_{ab}^x (f_{cd}^x)^* \frac{[12]^2 \ang{34}^2}{s_{14}s_{13}},
\eeq
while the $(\tilde{1}3)$ channel gives
\beq
-  \sum_x  f_{xa}^c(f_{xd}^b)^* \frac{[12]^2 \ang{34}^2}{s_{14}s_{12}}.
\eeq
Summing both, we get
\beq
\amp^{(1-4)}(1_a^+,2_b^+,3_c^-,4_d^-) =  - \sum_x \left(f_{ab}^x (f_{cd}^x)^* \frac{1}{s_{13}}+ f_{xa}^c(f_{xd}^b)^*\frac{1}{s_{12}} \right)   \frac{[12]^2 \ang{34}^2}{s_{14}}.  \label{amplitude14}
\eeq
If now we repeat the calculation with the (2-3) shift
\beq
\bra{\tilde{2}}= \bar{1} +z \bra{3},~~  |\tilde{3}] = |3] - z |2], ~~z \in \mathbb{C},
\eeq
we find instead
\beq
\amp^{(2-3)}(1_a^+,2_b^+,3_c^-,4_d^-) =  - \sum_x \left(f_{ab}^x (f_{cd}^x)^* \frac{1}{s_{13}}+ (f_{xc}^a)^* f_{xb}^d \frac{1}{s_{12}} \right)   \frac{[12]^2 \ang{34}^2}{s_{14}}. \label{amplitude23}
\eeq
The reason for the different prefactor in the second term is that the polarization with non-vanishing contribution in the $(1\tilde{3})$ channel is the opposite of the one that contributes in the $(\tilde{1}3)$ channel of the calculation with (1-4) shift. Requiring $\amp^{(1-4)}=\amp^{(2-3)}$, we find again the necessary condition $\sum_x f_{xa}^c(f_{xd}^b)^* = \sum_x (f_{xc}^a)^* f_{xb}^d$, which is the NAC in~\refeq{NAC}. The same condition is found by comparing the amplitudes $\amp^{(1-3)}$ and $\amp^{(2-4)}$, calculated with the shifts indicated in the superscripts.

\subsection{Modified Jacobi identities}
\label{sec:MJI}

Constructibility actually requires $\amp^{(i-j)}=\amp^{(i^\prime-j^\prime)}$ for any $i,i^\prime\in\{1,2\}$, $j,j^\prime \in \{3,4\}$, where again the superscripts indicate the (valid) shifts used to compute $\amp(1_a^+,2_b^+,3_c^-,4_d^-)$. Benincasa and Cachazo showed in~\cite{Benincasa:2007xk} that, under the assumptions that $\bar{f}=-f$ (so $f$ is imaginary) and $f$ is totally antisymmetric, the equalities with $i^\prime=i$ or with $j^\prime=j$ imply that these coupling constants obey the Jacobi identity. Without these assumptions, one gets instead  a different identity from each of the four choices, namely $\amp^{(1-3)}=\amp^{(1-4)}$, $\amp^{(1-3)}=\amp^{(2-3)}$, $\amp^{(1-4)}=\amp^{(2-4)}$ and $\amp^{(2-3)}=\amp^{(2-4)}$. The same four identities can be obtained by requiring the consistent factorization of $\amp(1_a^+,2_b^+,3_c^-,4_d^-)$ in different channels, as in~\cite{Schuster:2008nh,McGady:2013sga}, with each identity corresponding to a choice of complex factorized kinematics. At any rate, using NAC in~\refeq{NAC}, all these identities can be written as the modified Jacobi identity~\refeq{eq:MJI-1} (which explains the term NAC).
As emphasized in the introduction, this cannot be directly interpreted as the Jacobi identity for the structure constants of some Lie algebra. Note that~\refeq{eq:MJI-1} is invariant under unitary transformations in color space of the 1-particle states of the gluons. In terms of the matrices $F_a$ introduced above, the modified Jacobi identity can be written as
\beq
\left[F_a^\dagger,F_b \right] = \sum_c \left(f^b_{ac}\right)^* F_c. \label{MJImatrix}
\eeq
In this form, it is apparent that the $f$'s would form a Lie algebra if the matrices $F_a$ were Hermitian.

We have used NAC to reduce the four different possible modified Jacobi identities to a single one. Let us point out that, in fact, any of these versions of the modified Jacobi identity implies NAC. From the version in~\refeq{MJImatrix}, this can be shown easily by combining this equation with the one obtained by taking its adjoint.

\subsection{Real bases and the Jacobi identity }
\label{sec:hermiticity}

The NAC  derived above might at first sight seem of little
importance, but we shall now prove that it implies the existence of
a basis where the $F_{a}$ are Hermitian. 

We may proceed by considering the Hermitian matrix $N$ with entries
\begin{equation}
N_{\;\;b}^{a}:=\left\langle F_{a},F_{b}\right\rangle :=\textrm{tr}\left(F_{a}^{\dagger}F_{b}\right)
\end{equation}
and the symmetric matrix $K$ defined as
\begin{equation}
K_{ab}:=\left\langle F_{a}^{\dagger},F_{b}\right\rangle =\textrm{tr}\left(F_{a}F_{b}\right)\,.
\end{equation}
Using NAC, we obtain the relation
\begin{equation}
\sum_{b}\underbrace{\textrm{tr}\left(F_{a}F_{b}\right)}_{K_{ab}}\underbrace{\textrm{tr}\left(F_{b}^{\dagger}F_{c}\right)}_{N_{\;\;c}^{b}}=\sum_{b}\underbrace{\textrm{tr}\left(F_{a}F_{b}^{\dagger}\right)}_{\left(N_{\;\;b}^{a}\right)^{*}}\underbrace{\textrm{tr}\left(F_{b}F_{c}\right)}_{K_{bc}}
\end{equation}
which can be written simply as
\begin{equation}
KN=N^{*}K\,.\label{eq:KN-1}
\end{equation}
On the other hand
\begin{equation}
\sum_{b}\underbrace{\textrm{tr}\left(F_{a}F_{b}\right)}_{K_{ab}}\underbrace{\textrm{tr}\left(F_{b}^{\dagger}F_{c}^{\dagger}\right)}_{K_{bc}^{*}}=\sum_{b}\underbrace{\textrm{tr}\left(F_{a}F_{b}^{\dagger}\right)}_{\left(N_{\;\;b}^{a}\right)^{*}}\underbrace{\textrm{tr}\left(F_{b}F_{c}^{\dagger}\right)}_{\left(N_{\;\;c}^{b}\right)^{*}},
\end{equation}
i.e.
\begin{equation}
KK^{*}=N^{*}N^{*}\,.\label{eq:KK-1}
\end{equation}
To proceed, let us work in a basis in which $N$ is diagonal. This
can be achieved by a unitary redefinition $U_{1}$ of the particle
species:
\begin{equation}
N_{\;\;b}^{a}\rightarrow\left(U_{1}^{\dagger}NU_{1}\right)_{\;\;b}^{a}\overset{\cdot}{=}n_{a}\delta_{b}^{a}\,,n_{a}\in\mathbb{R}_{\geq0}\,.\label{eq:N-diag-1}
\end{equation}
Here and in the following a dot above the equal symbol indicates that the equality only holds in some bases. Of course, some degree of covariance is lost in such identities.
We note that the components $a$ associated with zero eigenvalues
of this matrix, $n_{a}=0$, correspond to $F_{a}=0$ and thus $f_{a\cdot}^{\cdot}=f_{\cdot a}^{\cdot}=0$.
Furthermore, from~(\ref{NAC}) we also conclude
that $f_{\cdot\cdot}^{a}=0$. Thus, the analysis of these components
is trivial, since they decouple completely from the remaining components:\footnote{We will denote the null subspace of $\mathcal{H}^+$ corresponding to the directions with $n_a=0$ by $\mathcal{H}_0=\mathrm{Ker}(N)$, and the subspace spanned by the eigenvectors of $N$ with non-vanishing eigenvalues, $n_a\neq 0$, by $\mathcal{H}_>$. Then we have $\mathcal{H}^+=\mathcal{H}_0 \oplus \mathcal{H}_>$.\label{footnote:H0etc}}
\begin{equation}
\left(n_{a}=0\right)\Leftrightarrow\left(f_{a\cdot}^{\cdot}=f_{\cdot a}^{\cdot}=f_{\cdot\cdot}^{a}=0\right)\,.
\label{eq:fAb=0}
\end{equation}
In the basis with diagonal $N$, equation (\ref{eq:KN-1}) imposes the constraint $\left(n_{a}-n_{b}\right)K_{ab}=0$.
Therefore $K$ must be block-diagonal, with blocks determined by the
degenerate eigenvalues of $N$ (i.e., those indices $a$ with the
same value of $n_{a}$ belong to the same block). We can then further
redefine the particle species inside each block so that $K$ becomes
fully diagonal, without spoiling (\ref{eq:N-diag-1}):
\begin{equation}
K_{ab}\rightarrow\left(U_{2}^{T}KU_{2}\right)_{ab}\overset{\cdot}{=}k_{a}\delta_{ab}\,,k_{a}\in\mathbb{R}_{\geq0}\,.
\end{equation}
Note that we choose the diagonal entries of this matrix to be real
and non-negative by absorbing any phases with $U_{2}$. Furthermore,
we observe that the possibility of simultaneously diagonalizing $K$
and $N$ is not too surprising, given that we made $N$ real, and
thus equation (\ref{eq:KN-1}) reduces to $\left[K,N\right]=0$.

Lastly, in this basis equation (\ref{eq:KK-1}) implies that $k_{a}=n_{a}$,
i.e.,
\begin{equation}
K\overset{\cdot}{=}N\,.
\end{equation}
But then
\begin{equation}
\left\Vert F_{a}^{\dagger}-F_{a}\right\Vert ^{2}=\left\langle F_{a}^{\dagger}-F_{a},F_{a}^{\dagger}-F_{a}\right\rangle =\underbrace{\left\langle F_{a}^{\dagger},F_{a}^{\dagger}\right\rangle }_{\left(K_{aa}\right)^{*}}+\underbrace{\left\langle F_{a},F_{a}\right\rangle }_{K_{aa}}-\underbrace{\left\langle F_{a}^{\dagger},F_{a}\right\rangle }_{N_{\;\;a}^{a}}-\underbrace{\left\langle F_{a},F_{a}^{\dagger}\right\rangle }_{N_{\;\;a}^{a}}\overset{\cdot}{=}0
\end{equation}
and so all $F_{a}$ must be Hermitian in such a basis, 
\beq
F_a^\dagger \doteq F_a, \label{hermitian}
\eeq
which is the same as
\begin{equation}
f_{ab}^{c}\overset{\cdot}{=}\left(f_{ac}^{b}\right)^{*}\,.
\end{equation}
Any orthonormal basis in which the $F_a$ are hermitian will be called hereafter a {\em real basis}.\footnote{This definition will be generalized when we introduce particles with other spins.} Working in a real basis, we can use~\refeq{hermitian} in~\refeq{MJImatrix}, which leads to the standard Jacobi identity,
\beq
\left[F_{a},F_{b}\right]=\sum_{c}f_{ab}^{c}F_{c}, \label{Jacobi}
\eeq
which can also be written as~\refeq{eq:JI}.
We have proved this in a specific basis. However the identity
is manifestly invariant under arbitrary unitary redefinitions of the
gluons in color space, so it actually holds in any basis, just like 
the modified Jacobi identity in~(\ref{MJImatrix}).
Therefore, without any extra assumption, such as CP symmetry, one can show that the self-couplings
of spin 1 massless particles do indeed exhibit the structure of a Lie algebra, with the couplings $f_{ab}^{c}$ playing the role of structure constants.

Furthermore, in a real basis, where $F_{a}\doteq F_{a}^{\dagger}$, by
conjugating the Jacobi identity we get that
\begin{equation}
\sum_{c}\left[f_{ab}^{c}+\left(f_{ab}^{c}\right)^{*}\right]F_{c}\overset{\cdot}{=}0\,.
\end{equation}
The $F_{c}$ are either null (when associated with a zero eigenvalue
of the $N$ matrix), in which case $f_{ab}^{c}$ is also null, or non-null and linearly independent. Either way we conclude
from the above constraint that in a real basis the $f_{ab}^{c}$
are purely imaginary, and so the tensor $f$ is fully antisymmetric.

The total antisymmetry of the $f$ tensor is a basis-dependent property.
However, we observe that
\begin{equation}
\textrm{tr}\left(\left[F_{a},F_{b}\right]F_{c}\right)=\sum_{d}f_{ab}^{d}K_{dc}\overset{\cdot}{=}\sum_{c}f_{ab}^{c}n_{c}\,,
\end{equation}
where the last equality applies in the real basis with diagonal $K$, introduced above. Since the trace on the left is always fully antisymmetric in $abc$,
it follows that the tensor $\sum_{d}f_{ab}^{d}K_{dc}$ shares the
same property (and as a consequence, in our special basis, for every
non-null entry $f_{ab}^{c}$ one must have $n_{a}=n_{b}=n_{c}$).
Likewise, the $F_{a}$ are not Hermitian in a general basis. However, starting with~\refeq{hermitian} and performing a general unitary change of basis, it is clear that the $F_{a}^{\dagger}$ are always a linear combination of the $F_{a}$. More explicitly,
by applying NAC to $\sum_{b}F_{b}^{\dagger}\textrm{tr}\left(F_{b}F_{a}\right)$
one finds the relation
\begin{equation}
\sum_{b}K_{ab}F_{b}^{\dagger}=\sum_{b}N_{\;\;a}^{b}F_{b} .\label{KFdNF}
\end{equation}
Hence, in an arbitrary orthonormal basis the hermiticity condition translates into 
\beq
F_{a}^{\dagger}=\sum_{b}\Omega^{ab}F_{b},
\eeq
for $\Omega=K^{\mathrm{MP}}N^{*}$, 
where $K^{\mathrm{MP}}$ is the Moore-Penrose pseudoinverse of $K$.

In more formal terms, we have shown that a set of coupling constants $f^c_{ab}$ satisfying the NAC condition defines a {\it real structure} in the complex space ${\cal H}_>$ (introduced in footnote~\ref{footnote:H0etc}). That is, it singles out a real subspace, ${\cal H}_{\mathbb{R}}\subset {\cal H}_>$, consisting of the real linear combinations of the elements of a real basis. Consequently, ${\cal H}_>={\cal H}_{\mathbb{R}}\oplus i{\cal H}_{\mathbb{R}}$. Notice that the real structure is only defined on ${\cal H}_>$ with this definition, because the condition of hermitian generators is trivial in the null subspace $\mathcal{H}_0$. A more general real structure will be defined in the next section when considering the interactions of gluons with other particles. The matrix $\Omega$ introduced before can be seen as a linear isometry between $\mathcal{H}_>^*$ and $\mathcal{H}_>$ that characterizes the real structure on $\mathcal{H}_>$. Indeed, composing it with the antilinear isometry between these two spaces, it provides an antiunitary involution $\tilde{\Omega}: \mathcal{H}_>\to\mathcal{H}_>$, which corresponds to the conjugation under this real structure. 

\section{Interactions of spin 1 particles with other particles}
\label{sec:matter}

In the presence of particles with other spins, there is a similar
concern that consistent factorization of four-point amplitudes might
not actually impose that the new couplings have the structure of a
Lie algebra. We consider minimal couplings of gluons with other massless particles, that is, the couplings of massless spin 1 particles to pairs of particles with opposite helicity. Analogously to the situation for gluon self-couplings, the extra ingredient we need to infer a Lie algebra is the existence of a basis in which these coupling constants, when written as matrices, are Hermitian. A non-trivial aspect of this is that we need hermiticity to hold for all couplings in the same basis.

In the following, we show that such a basis does exist for the coupling of gluons to all pairs
of particles with opposite helicities (scalars, spin 1/2 fermions and spin~1 bosons): there is a basis where all these matrices are
simultaneously Hermitian. We work in connected particle bases\footnote{See discussion in section~\ref{sec:CPT}.} such that $\mathrm{CPT}\left|\vec{p}_{j}^{\,h}\right\rangle =(-1)^{\left|h\right|-h}\left|\vec{p}_{j}^{\,-h}\right\rangle$.\footnote{Note that in the case of scalars, $h=0$, we work in a basis that is self-conjugate under CPT. Thus, the basis transformations of scalars must be orthogonal and upper and lower indices transform in the same way.} Similarly to our method in section~\ref{sec:NAC}, we consider the factorization
$1_{i}^{h}2_{j}^{-h}\rightarrow X_{x}^{\pm m}\rightarrow3_{k}^{h^{\prime}}4_{l}^{-h^{\prime}}$
of the four-particle amplitude $\mathcal{A}\left(1_{i}^{h},2_{j}^{-h},3_{k}^{h^{\prime}},4_{l}^{-h^{\prime}}\right)$
mediated by helicity $\pm m$ massless bosons, with $m\in\{1,2\}$ and $|h|,|h^\prime|\leq m$, to avoid unphysical poles. Little-group covariance requires, for $h,h^{\prime}\geq0$,
\beq
\amp \left(1_{i}^{h},2_{j}^{-h},3_{k}^{h^{\prime}},4_{l}^{-h^{\prime}}\right) =
[13]^{h+h^\prime}\ang{24}^{h+h^\prime} [14]^{h-h^\prime} \ang{23}^{h-h^\prime} H_{ijkl}(s_{12},s_{13}),\label{generalA4}
\eeq
for some flavor and momentum-dependent function $H$.
We consider cubic minimal couplings of the intermediate boson, with coupling constants defined by
\begin{align}
\mathcal{A}\left(1_{a}^{m},2_{i}^{h},3_{j}^{-h}\right) & = i^{2h}C_{ai}^{(h,m)\,j} \left(\frac{\left[12\right]\left[31\right]}{\left[23\right]}\right)^{m}\left(\frac{\left[12\right]}{        \left[31\right]}\right)^{2h}
\label{generalcubic+} \\
\mathcal{A}\left(1_{a}^{-m},2_{i}^{-h},3_{j}^{h}\right) & =
i^{-2h}\bar{C}_{j}^{(h,m)\,ai}\left(\frac{\left\langle 12\right\rangle \left\langle 31\right\rangle }{\left\langle 32\right\rangle }\right)^{m}\left(\frac{\left\langle 12\right\rangle }{\left\langle 31\right\rangle }\right)^{2h}\,, \label{generalcubic-}
\end{align}
with $h\geq 0$ and $m\in\{1,2\}$. The CPT relation implies $\bar{C}_j^{(h,m)\,ai} = \left(C_{ai}^{(h,m)\,j} \right)^*$.
The $s_{12}$-channel factorization of the
four-particle amplitude $\mathcal{A}\left(1_{i}^{h},2_{j}^{-h},3_{k}^{h^{\prime}},4_{l}^{-h^{\prime}}\right)$, with $h,h^{\prime}\geq0$,
can follow two branches, according to the complex kinematics~\cite{Schuster:2008nh}. Let us call $P_{-}$ the residue when $[12]$ vanishes and $P_{+}$ the residue when $\ang{12}$ vanishes. They must take the values 
\begin{align}
P_{-} & =\sum_{x} (-1)^{2h} \mathcal{A}\left(X_{x}^{-m},2_{j}^{-h},1_{i}^{h}\right)\mathcal{A}\left(-X_{x}^{m},3_{k}^{h^{\prime}},4_{l}^{-h^\prime}\right)\nonumber \\
 & =\sum_{x}i^{2\left(h^{\prime}-h\right)}\left(C_{xj}^{(h,m)\,i}\right)^*C_{xk}^{(h^{\prime},m)\, l}K_{h,m}(X,2,1)^* K_{h^{\prime},m}(-X,3,4)\,,\\
P_{+} & = (-1)^{2h^\prime} \sum_{x}\mathcal{A}\left(X_{x}^{m},1_{i}^{h},2_{j}^{-h}\right)\mathcal{A}\left(-X_{x}^{-m},4_{l}^{-h^{\prime}},3_{k}^{h^\prime}\right)\nonumber \\
 & =\sum_{x}i^{2\left(h-h^{\prime}\right)}C_{xi}^{(h,m)\,j}\left(C_{xl}^{(h^{\prime},m)\,k}\right)^*K_{h,m}(X,1,2) K_{h^{\prime},m}(-X,4,3)^*\, ,
\end{align}
where 
\beq
K_{h,m}(1,2,3) = \left(\frac{\left[31\right]\left[12\right]}{\left[23\right]}\right)^{m}\left(\frac{\left[12\right]}{\left[31\right]}\right)^{2h} ,
\eeq
and we have taken spin-statistics into account.\footnote{The standard spin-statistics relation can be derived from the usual S-matrix postulates~\cite{Stapp:1962nxd,Weinberg:1964cn,Arkani-Hamed:2017jhn}.} 
Consistency of~\refeq{generalA4} with both choices of factorization kinematics requires $P_{+}=P_{-}$. It can be checked explicitly that, for $s_{12}=0$,
\begin{equation}
\left(-1\right)^{2h^{\prime}}K_{h,m}(X,2,1)^* K_{h^{\prime},m}(-X,3,4) =\left(-1\right)^{2h}K_{h,m}(X,1,2) K_{h^{\prime},m}(-X,4,3)^*\,,
\end{equation}
so the dependence on kinematics is indeed the same. For complete agreement we need
\begin{equation}
\sum_{x}\left(C_{xj}^{(h,m)\,i}\right)^* C_{xk}^{(h^{\prime},m)\,l}=\sum_{x}C_{xi}^{(h,m)\,j} \left(C_{xl}^{(h^{\prime},m)\,k}\right)^* \label{pregeneralNAC}
\end{equation}
Defining the matrices $C_a^{(h,m)}$ with entries 
\beq
\left(C^{(h,m)}_a\right)^i_{~j} = C_{aj}^{(h,m)\,i},
\eeq
\refeq{pregeneralNAC} reads
\begin{equation}
\sum_{x}\left[C_{x}^{(h,m)}\right]^{\dagger}\otimes C_{x}^{(h^{\prime},m)}=\sum_{x}C_{x}^{(h,m)}\otimes\left[C_{x}^{(h^{\prime},m)}\right]^{\dagger} \label{generalNAC}
\end{equation}
This is a generalization of the NAC in~\refeq{NAC} for any combination
of helicities $\left(h,h^{\prime}\right)$, and is valid not only
for minimal couplings of spin $m=1$ particles but also for those of gravitons
($m=2$). Focusing on the interactions of gluons, let us call $\Theta_{a}=C^{(0,1)}_a$, $T_a = C^{(1/2,1)}_a$, $F_a=C^{(1,1)}_a$. 
These matrices encode the minimal couplings of gluons to scalars, spin 1/2 fermions and self-couplings.
The condition we just derived can be written as
\begin{equation}
\sum_{a}X_{a}\otimes X_{a}^{\dagger}=\sum_{a}X_{a}^{\dagger}\otimes X_{a} \label{XNAC}
\end{equation}
for the block-diagonal matrices
\begin{equation}
X_{a}:=\left(\begin{array}{ccc}
\Theta_{a} & 0 & 0\\
0 & T_{a} & 0\\
0 & 0 & F_{a}
\end{array}\right)\,.
\end{equation}
Then the argument in subsection~\ref{sec:hermiticity} can be repeated, using instead of $N$ and $K$ the matrices $N^{(X)}$ and $K^{(X)}$ with entries
\beq
N^{(X)a}_b := \mathrm{tr}\left(X^\dagger_a X_b\right),~~K^{(X)}_{ab} := \mathrm{tr}\left(X_a X_b\right),\label{eq:NXKX}
\eeq
to show that there is a basis where the $X_{a}$ are Hermitian. Given
their block-diagonal form, this in turn implies that it is possible
to find a basis where all $\Theta_{a}$, $T_{a}$ and $F_{a}$ are {\em
simultaneously\/} Hermitian. Hereafter a {\em real basis} is defined as one in which $X_a\doteq X_a^\dagger$ or, equivalently, $N^{(X)}\doteq K^{(X)}$. This generalizes our previous definition for gluons and the corresponding real structure, which is extended in a specific way to the subspace of $\mathcal{H}_0$ with non vanishing $X_a$.

From the arguments in~\cite{Benincasa:2007xk,Schuster:2008nh,McGady:2013sga}, it follows that consistent factorization in all channels
of $\mathcal{A}\left(1_{a}^{h},2_{b}^{-h},3_{c}^{h^{\prime}},4_{d}^{-h^{\prime}}\right)$
requires the modified algebra relation
\begin{align}
\left[X_{a}^{\dagger},X_{b}\right] & =\sum_{x}f_{bx}^{a}X_{x}^{\dagger} .\label{modifiedalgebra}
\end{align} 
This implies that
\begin{align}
\left[X_{a},X_{b}\right] & =\sum_{x}f_{ab}^{x}X_{x} \label{generalAlgebra}
\end{align}
in any real basis. In fact, being covariant, this relation holds as well in any basis. Let us also observe that the third row of the matrix equation~\refeq{XNAC}, 
\begin{equation}
\sum_{a}X_{a}\otimes F_{a}^{\dagger}=\sum_{a}X_{a}^{\dagger} \otimes  F_{a},
\end{equation} 
can be derived from~\refeq{modifiedalgebra}, by combining this equation with its adjoint. 
While, using the results of the previous section, this implies that $X_a$ is Hermitian in a real basis for $a$ in $\mathcal{H}_>$, this alone is not sufficient to prove the existence of a real basis in which $X_a$ are Hermitian for the gluon states $a$ in~$\mathcal{H}_0$.

\section{Yang-Mills symmetry}
\label{sec:symmetry}

We have shown that the modified algebra relations, together with the existence of real bases, imply that the coupling constants of cubic amplitudes with at least one gluon form a complex Lie algebra $\mathcal{L}$. The commutation relations can be written as~\refeq{generalAlgebra}
in any orthonormal basis, so these coupling constants are the matrices of a representation of the algebra and the three-gluon coupling constants $f_{ab}^c$ are the structure constants. Furthermore, in real bases these structure constants are purely imaginary and completely antisymmetric. It is important to stress that this in turn implies (see, for instance,~\cite{Weinberg:1996kr}) that $\mathcal{L}$  is necessarily a reductive Lie algebra, that is, it is the direct sum of a semisimple algebra (with vector space $\mathcal{H}_>$) and an abelian one (with vector space $\mathcal{H}_0)$. Therefore, the Lie algebras that are relevant for theories with massless spin 1 particles belong to a quite particular class. Generic Lie algebras,\footnote{Most Lie algebras are not reductive. In fact, for dimension~$\geq2$, the space of structure constants  corresponding to reductive algebras is a proper subvariety, so it has measure zero.} such as the 2-dimensional algebra mentioned in the introduction, are excluded; they are not consistent with the physical principles of Poincar\'e symmetry, unitarity and locality. 

Let us now define the associated symmetry transformations. We start in a real basis and define infinitesimal linear transformations of a particle $\ket{j}$ of helicity $h_j$ as $\delta(\xi) \ket{j} := \sum_a \xi_a \delta_a \ket{j}$, with
\beq
\delta_a \ket{j} := \begin{cases}
i \sum_k (X_a)_j^{~k} \ket{k}, ~~ h_j\geq 0 , \\
-i \sum_k (X_a^*)^j_{~k} \ket{k}, ~~ h_j < 0   .
\end{cases} 
\label{transformation} 
\eeq
Note that $X_a$ transforms particles into particles with the same helicity (and momentum). Note also that antiparticles transform in conjugate representations and that the representations are real in the blocks with integer helicity. The parameters $\xi_a$ must be real, in order to preserve the Hilbert inner product. Then the generators written in the transformed basis will still be Hermitian. Moreover, for real $\xi_a$, $\delta(\xi) \ket{j} = i X(\xi) \ket{j}$, where $X(\xi)$ is a representation of a particular real form $\mathcal{L}_\mathbb{R}$ of the complex algebra $\mathcal{L}$: the one formed by real linear combinations of the Hermitian generators, that is, vectors in~$\mathcal{H}_\mathbb{R}$.  Because the complex Lie algebra is reductive, this real form is the direct sum of a compact real Lie algebra and an abelian one. Exponentiation leads to a compact Lie group times an abelian one (which could also be compact).\footnote{Our perturbative analysis gives no topological information.}
Under these transformations, each coupling constant in an amplitude transforms as a tensor according to its index structure.\footnote{In an arbitrary orthonormal basis, the antisymmetry of the structure constants takes the form
\begin{equation}
\sum_x \left( \Omega^\dagger_{xa} f^x_{bc} + \Omega^\dagger_{bx} f^x_{ac} \right) =0,
\end{equation}
which shows that $\Omega$ is an invariant tensor under the action of the Lie group. This is the covariant formulation of the reality of the Lie algebra.}

The relevance of these transformations is that they leave all amplitudes invariant. For cubic amplitudes involving at least one spin 1 particle, this  follows in a real basis from the algebra relations: 
\begin{align}
-i \delta_a (X_b)_i^{~j} & = \sum_c (X_a)_b^{~c}  (X_c)_i^{~j} +  \sum_k (X_a)_i^{~k} (X_b)_k^{~j} - \sum_k (X_a^*)^j_{~k} (X_b)_i^{~k}  \nn
& \overset{\cdot}{=}  \sum_c f^c_{ba}  (X_c)_i^{~j} +  \sum_k (X_a)_i^{~k} (X_b)_k^{~j} - \sum_k (X_b)_i^{~k}   (X_a)_k^{~j}  \nn
& = \left(- \sum_c f^c_{ab} X_c + X_a X_b - X_b X_a \right)_i^{\,j} \nn
& = 0.
\end{align}
Since the metric that enters in the factorization equations corresponds to the Hilbert inner product, which is invariant, any amplitude constructed with these cubic amplitudes, to any order in loops, will also be invariant.

Moreover, the very presence of massless spin 1 particles implies the invariance under this transformation of arbitrary amplitudes, even when they do not involve spin 1 particles. This has been proven explicitly with on-shell methods in a few examples, such as local cubic interactions of massive scalars~\cite{Arkani-Hamed:2017jhn}. More generally, it follows from Weinberg's soft theorem for spin 1 massless particles~\cite{Weinberg:1965nx}, which has been also been proven with purely on-shell methods in~\cite{Elvang:2016qvq,Falkowski:2020aso}.
{\em A priori,} there are two independent spin-1 soft theorems: one for helicity +1 and one for helicity -1. The modified Jacobi identity~\refeq{MJImatrix} and the modified algebra relation~\refeq{modifiedalgebra} actually follow directly by applying the +1 soft theorem to the cubic amplitude $\amp(1^{-1}_a,2^{-h},3^{+h})$, or equivalently the -1 soft theorem to $\amp(1^{1}_a,2^{h},3^{-h})$.\footnote{Applying the $\pm 1$ soft theorem to $\amp(1^{\pm 1}_a,2^{\pm h},3^{\mp h})$ just shows that the amplitudes $\amp(\pm1,\pm1,\pm h,\mp h)$ (with minimal couplings) vanish at tree level.} Our results above then show {\em a posteriori\/} that the two types of soft theorem are actually one and the same. Note that the invariance of amplitudes applies also to those involving non-renormalizable interactions, such as the one for three gluons with the same helicity.

We conclude that the transformations $\delta(\xi)$ defined by ~\refeq{transformation} (and their finite versions) are a non-trivial symmetry of the S matrix in any theory with interacting spin 1 particles.

\section{CP symmetry}
\label{sec:CP}

In a general orthonormal basis, we can define the CP transformation of 1-particle states for massless particles with helicity $\pm h$ ($h\geq 0$) as the unitary transformation~\cite{Branco:1999fs} 

\begin{align}
\mathrm{CP} \ket{\vec{p}_f^{\,h}} & =  \sum_{f^{\prime}} R^{(h)}_{f f^\prime}\,\ket{-\vec{p}_{f^\prime}^{\,-h}}, \label{CP+} \\
\mathrm{CP} \ket{\vec{p}_f^{\,-h}} & = \sum_{f^{\prime}} \bar{R}^{(h)\,f f^\prime}\,\ket{-\vec{p}_{f^\prime}^{\,h}}, \label{CP-}
\end{align}
where $R^{(h)}$ and $\bar{R}^{(h)}$ are unitary matrices that only mix particles with the same helicity modulus. Note that $\bar{R}^{(0)} = R^{(0)}$. From the CPT isometry we have the relation 
\beq
\bar{R}^{(h)\,f f^\prime} = \left(R^{(h)}_{f f^\prime} \right)^* .
\eeq
In particular, $R^{(0)}$ is a real orthogonal matrix.\footnote{This arises from our CPT identification of spin 0 particles as their own antiparticles. In field theory, this corresponds to working with real fields.}

Now, we have shown above that the minimal gluon couplings $X_a$ are representations of a reductive Lie algebra. Then, the conditions that CP transformations must fulfill in order to leave these couplings invariant are exactly the same as the ones imposed in~\cite{Grimus:1995zi} from field-theoretical considerations. Therefore, the algebraic results in~\cite{Grimus:1995zi} can be directly applied and we conclude that, in any theory with gluons, there exist (theory and basis-dependent) unitary matrices $R^{(h)}$ such that the CP transformation in~\refeq{CP+} and~\refeq{CP-} leaves the minimal interactions of gluons invariant. 

Using the CPT relation as before, which dresses CP in the guise of T, this translates into the following property of the couplings:
\begin{equation}
X_{a}=-\sum_{b} R_{a b}^{(1)}\left(R^{\dagger}X_{b}^{*}R\right)\,,~~~~~~~R:=\left(\begin{array}{ccc}
R^{(0)} & 0 & 0\\
0 & R^{(1/2)} & 0\\
0 & 0 & R^{(1)}
\end{array}\right)\,.\label{eq:CP-X}
\end{equation}

Let us stress that, working in a particle basis that is real and self-conjugated for the $h=\pm 1$ and $h=0$ sectors respectively, it is not possible in general to choose matrices $R^{(h)}$ for which the invariance condition is fulfilled and such that they are all the identity.
This would be true, however, in the absence of fermions. Indeed, working in a real basis and choosing trivial matrices $R^{(0,1)}$, eq.~\refeq{eq:CP-X} simply tells us that the couplings $F_a$ and $\theta_a$ are purely imaginary, which we had already proven (in that basis) above. However, the fermion generators $T_a$ are, in general, complex representations, and their complex conjugation has to be compensated, in general, by a non-trivial $R^{(1)}$. Then, CP invariance of the $F_a$ requires $R^{(1)}$ to be an automorphism of the algebra. And finally, a non-trivial $R^{(1)}$ must be compensated by a non-trivial $R^{(0)}$ in the scalar sector. That all this is possible is non-trivial and we refer to~\cite{Grimus:1995zi} for details. Note finally that this proof of CP invariance applies only to the (minimal) gluon couplings and only at the tree level. It cannot be extended, in particular, to Yukawa couplings and scalar self-interactions. For those interactions, the requirement of CP invariance imposes non-trivial constraints.

\section{Gravity}
\label{sec:gravity}

Let us consider finally massless particles of helicity $\pm 2$, called gravitons in the following, with pairs of antiparticles identified as above. Their leading cubic amplitudes are given in general by 
\begin{align}
& \mathcal{A}\left(1_{a}^{+2},2_{b}^{+2},3_{c}^{-2}\right) = g_{ab}^c \frac{[12]^6}{[13]^2[23]^2}, \\
& \mathcal{A}\left(1_{a}^{-2},2_{b}^{-2},3_{c}^{+2}\right) = {\bar{g}}^{ab}_c \frac{\ang{12}^6}{\ang{13}^2\ang{23}^2}, 
\end{align}
where $g$ and $\bar{g}$ are flavor-dependent complex tensors of the type specified by their indices and with mass dimension $-1$. The CPT relation implies $\bar{g}^{ab}_c = \left(g_{ab}^c\right)^*$. The kinematics and Bose symmetry require that $g$ ($\bar{g}$) be symmetric under the exchange of the lower (upper) indices. On the other hand, the symmetry under the exchange of lower and upper indices is not guaranteed at this level. This property can only be basis dependent, at best. The same applies to the reality of $g$, which cannot be inferred from cubic interactions without the assumption of CP symmetry of graviton interactions. And anyway, the reality of $g$ does not imply its complete symmetry.

From the consistent factorization of four-graviton amplitudes it has been proven in~\cite{Benincasa:2007xk} (see also \cite{Schuster:2008nh,McGady:2013sga,Arkani-Hamed:2017jhn}) that there is a basis in which all the gravitons decouple from each other, as shown in~\cite{Boulanger:2000rq} with BRST methods. 
This proof assumes the complete symmetry of $g$, which, as we have just argued, does not follow from cubic interactions. However, it should be already clear that four-graviton amplitudes contain the required missing information. Explicitly, consistent factorization requires 
\begin{align}
& \sum_x g^x_{ab} \bar{g}^{cd}_x = \sum_x g^d_{ax} \bar{g}^{cx}_b = \sum_x g^c_{ax} \bar{g}_b^{dx} , \label{factorization2} \\
& \sum_x g^c_{ax} \bar{g}^{dx}_b = \sum_x g^d_{bx} \bar{g}^{cx}_a  \label{NAC2}.
\end{align}
Eq.~\refeq{NAC2} is the NAC already derived in section~\ref{sec:matter} (for the choice $m=2$). It can also be obtained by combining the last equality in~\refeq{factorization2} with its complex conjugate. Using the matrix notation $(G_a)^c_b := g_{ab}^c$ and the CPT relation, the graviton NAC is saying that 
\beq
\sum_a G_a \otimes G_a^\dagger = \sum_a G_a^\dagger \otimes G_a.
\eeq
The steps taken for the $F$'s can be repeated here, leading to a basis where 
\beq
G_a^\dagger \doteq G_a .
\eeq
Graviton orthonormal bases with this property will also be called real bases. The changes of bases that preserve the reality property are orthogonal. Moreover, in matrix notation,~\refeq{factorization2} reads
\beq
G_a G_b^\dagger = G^\dagger_b G_a = \sum_x g^b_{ax} G_x^\dagger. \label{factorization2matrix}
\eeq
In a real basis, these equalities imply that $g^c_{ab}$ is real, and therefore fully symmetric, i.e.
\beq
G_a \doteq G^*_a \doteq G_a^T.
\label{realsymmetric}
\eeq
Furthermore, in such a basis the first equality in~\refeq{factorization2matrix} is $[G_a,G_b] \doteq 0$, so all these real symmetric matrices can be simultaneously diagonalized by orthogonal transformations. Note that an orthogonal change of basis will also act on the explicit index $a$ in $G_a$, but the resulting real linear combination of real diagonal matrices will also be diagonal and real, so  in the new real basis we can write
\beq
g_{ab}^c \doteq g(a,b) \delta_{bc} . 
\eeq
The orthogonal transformations keep the tensor $g$ fully symmetric, so
\beq
g_{ab}^c \doteq \begin{cases} g(a,b) \delta_{cb} \\ g(c,b) \delta_{ab}
\end{cases}~,
\eeq
and hence 
\beq
g_{ab}^c \doteq g_a \delta_{ab}\delta_{bc},
\eeq
which expresses the decoupling of different gravitons in this basis. The steps after~\refeq{realsymmetric} are just an explicit version of the argument in~\cite{Benincasa:2007xk}.

Consider now the minimal coupling of one graviton to two massless particles of opposite helicity, which is given by~\refeq{generalcubic+} and~\refeq{generalcubic-} with $m=2$. Consistent factorization requires~\cite{Benincasa:2007xk,Schuster:2008nh,McGady:2013sga} that the other massless particles coupling to the graviton in this amplitude have helicity $|h|\leq 2$. Let us define the block-diagonal matrices $Y_a$ with entries
\beq
\left( Y_a \right)_{(b,h)}^{(c,h^\prime)} := C_{ba}^{(h,2)\,c} \delta_{h h^\prime},~~~0 \leq h,h^\prime \leq 2.
\eeq
Note that the $G_a$ form the $h=2$ block of $Y_a$.  
In terms of these matrices, the NAC in~\refeq{generalNAC} reads
\beq
\sum_a Y_a \otimes Y_a^\dagger = \sum_a Y^\dagger_a \otimes Y_a .
\eeq
The same proof as for spin 1 implies that there is a {\em real} basis in which the $Y_a$ are Hermitian. 

Applying Weinberg's gravity soft theorem~\cite{Weinberg:1965nx}, proven in~\cite{Elvang:2016qvq,Falkowski:2020aso} with pure on-shell methods, to a cubic amplitude of massless particles involving a graviton with opposite helicity, we find
\beq
Y_a Y_b^\dagger = Y^\dagger_b Y_a = \sum_x g^{c}_{ax} Y^\dagger_x .
\eeq
This relation has been first found with on-shell methods in~\cite{Benincasa:2007xk}, in the case of diagonal and real $Y_a$. Eq.~\refeq{factorization2} is one of the blocks in these matrix identities. In the real basis with decoupling gravitons (in which the $h=2$ block is diagonal) this gives
\begin{align}
& [Y_a,Y_b]\doteq 0 ,\\
& Y_a Y_b \doteq  g_a \delta_{ab} Y_a .
\end{align}
The first  of these equations implies that the $Y_a$'s can be diagonalized simultaneously also in the blocks with $h\neq 2$. The second one shows that $\{ Y_a/g_a \}$ is a set of orthogonal projectors (with the exception of $Y_a$ for vanishing $g_a$, which are simply null matrices). Their eigenvalues are either 1 or 0, showing that, in the real basis with diagonal $Y_a$'s, the following holds: i) the non-vanishing couplings of any particle to graviton $a$ is $g_a$; ii) a given particle cannot couple to more than one species of graviton. The first condition is the universality of cubic graviton couplings, while the second one shows that no particle can connect directly different graviton species. 

One may wonder if different species of gravitons might still be connected by intermediate non-gravitational interactions of particles that can couple to different gravitons. However, this is also forbidden by Weinberg's soft theorem. Indeed, for any non-vanishing amplitude $\amp(ij\dots)$ we have, schematically,
\beq
\sum_x \left(Y_a\right)_i^x \amp(xj\dots) =  \sum_x \left(Y_a\right)_j^x \amp(ix\dots) .
\eeq
In the real basis with  diagonal $Y_a$ (with elements equal to 0 or $g_a$, as shown above), it follows that either both $i$ and $j$ couple to graviton $a$ (with strength $g_a$), or both $i$ and $j$ have vanishing coupling to graviton $a$. Therefore, any particle theory fulfilling the axioms of S matrix theory and with two or more massless particles of helicity $\pm 2$ is separated into completely independent sectors that do not interact with each other. This agrees with the field-theoretical results in~\cite{Boulanger:2000rq}.
Because the couplings $Y_a$ are essentially projectors, it is obvious that gravitational interactions are CP conserving, with the action of CP on gravitons defined in the straightforward way (without rotations) in each decoupled sector. Clearly, this does not affect the CP transformations of gluon interactions.

\section{Conclusions}
\label{sec:Conclusions}

The inevitability of Yang-Mills symmetry in a theory with massless spin 1 particles for the consistency of scattering amplitudes with fundamental physical principles is well known. The same applies to the decoupling of different species of massless spin 2 particles and the universality of their interactions. However, the standard proofs of these facts rely on symmetry properties of the cubic coupling constants that are not guaranteed by the form of three-particle amplitudes. Sometimes it is assumed that these interactions are parity invariant, but in fact this cannot be used to prove all the required symmetry properties.

In this work, we have completed these proofs by showing without extra assumptions that the consistent factorization of four-particle amplitudes contains the necessary missing piece of information: a non-ambiguity condition (NAC) from which one can derive the existence of a set of bases defining a real structure, in which the matrices defined from the coupling constants are Hermitian. Then, the usual on-shell arguments can be used to show that i) in those bases the cubic self-interactions of massless spin 1 and spin 2 particles obey the expected symmetry and reality properties; ii) the amplitudes are invariant under a Yang-Mills symmetry generated by the couplings of the massless spin 1 particles; iii) the corresponding Lie group is the product of a compact group and an abelian one; iv) the cubic couplings of massless spin 1 and 2 particles with any massless particle are CP invariant; and v) a theory with massless spin 2 particles is divided into completely decoupled sectors with one massless spin 2 particle each (the graviton), which couples universally to all the particles in that sector.

Once the existence of real bases is established, all the results above follow in fact from Weinberg's soft theorems~\cite{Weinberg:1965nx} for massless spin 1 and spin 2 particles, which can also be proven from consistent factorization, with Lorentz symmetry enforced from the start~\cite{Elvang:2016qvq,Falkowski:2020aso}. In a sense, the spin 1 particles act as generators of the Yang-Mills transformations, while the transformations induced by spin 2 particles do not mix different flavors and act universally in each sector.\footnote{On the other hand, the space-time transformation is non-trivial and the soft gravitons induce the Ward identities of the Bondi–Metzner–Sachs group~\cite{Bondi:1962px,Sachs:1962wk,Strominger:2017zoo}. Note that its Poincar\'e subgroup is an input in the on-shell formalism.}

There is also an analogous soft theorem for massless spin 3/2 particles, that is, for gravitinos~\cite{Grisaru:1977kk,Liu:2014vva,Avery:2015iix,Lysov:2015jrs}. Their cubic minimal couplings induce supersymmetry transformations and the soft theorem implies that a theory with interacting massless spin 3/2 particles must be supersymmetric. Unlike the case of spin 1 and 2 particles, no extra property of the cubic couplings is needed to prove these consequences of the existence of spin 3/2 particles. This is fortunate, as there is no gravitino analogue of NAC, since the helicity of an exchanged gravitino (or more generally, any exchanged fermion) is dictated by the helicities of the external particles it minimally couples to.  

Besides the tree-level constraints we have revisited here, one must enforce the cancellation of gauge anomalies. Of course, in the on-shell formalism, which avoids gauge redundancies, the notion of gauge anomalies does not make sense, but there are equivalent problems with chiral fermions, which show up as some potential inconsistency of factorization at the loop level. These anomalies, and the conditions for their cancellation, have been discussed and analyzed from an on-shell perspective in~\cite{Huang:2013vha,Chen:2014eva,Alviani:2025msx}.

Before finishing, let us briefly comment on the  couplings  of gluons and gravitons to massive particles. In this work, we have considered the minimal couplings of massless particles, which are the ones with the lowest dimensionality. In the massive case, minimal couplings can be defined as the ones that correspond to massless minimal coupling in the high-energy limit. For these couplings, we can simply apply our results for the massless case.

\acknowledgments

This work has been partially supported by grants PID2022-139466NB-C21 and PID2022-139466NB-C22, funded by MICIU/AEI/10.13039/501100011033 and by ERDF/EU A way of making Europe, by the Junta de Andalucía grants FQM 101 and  P21\_00199 (FEDER) and by Consejería de Universidad, Investigación e Innovación, 
Gobierno de España and Unión Europea – NextGenerationEU under grant $\mathrm{AST22\_6.5}$.
The work of C.H.G is further supported by the Spanish government under the FPI grant PREP2022-000831.
The work of J.M.L. has been supported by the grant CSIC-20223AT023. J.M.L. also acknowledges the support of the Spanish Agencia Estatal de Investigacion through the grant “IFT Centro de Excelencia Severo Ochoa CEX2020-001007-S”.

\bibliographystyle{JHEP}
\bibliography{references}

@book{Weinberg:1996kr,
    author = "Weinberg, Steven",
    title = "{The quantum theory of fields. Vol. 2: Modern applications}",
    doi = "10.1017/CBO9781139644174",
    isbn = "978-1-139-63247-8, 978-0-521-67054-8, 978-0-521-55002-4",
    publisher = "Cambridge University Press",
    month = "8",
    year = "2013"
}

@book{Branco:1999fs,
    author = "Branco, Gustavo C. and Lavoura, Luis and Silva, Joao P.",
    title = "{CP Violation}",
    doi = "10.1093/oso/9780198503996.001.0001",
    isbn = "978-1-383-02075-5, 978-0-19-850399-6",
    volume = "103",
    year = "1999"
}

@article{Britto:2004ap,
    author = "Britto, Ruth and Cachazo, Freddy and Feng, Bo",
    title = "{New recursion relations for tree amplitudes of gluons}",
    eprint = "hep-th/0412308",
    archivePrefix = "arXiv",
    doi = "10.1016/j.nuclphysb.2005.02.030",
    journal = "Nucl. Phys. B",
    volume = "715",
    pages = "499--522",
    year = "2005"
}

@article{Britto:2005fq,
    author = "Britto, Ruth and Cachazo, Freddy and Feng, Bo and Witten, Edward",
    title = "{Direct proof of tree-level recursion relation in Yang-Mills theory}",
    eprint = "hep-th/0501052",
    archivePrefix = "arXiv",
    doi = "10.1103/PhysRevLett.94.181602",
    journal = "Phys. Rev. Lett.",
    volume = "94",
    pages = "181602",
    year = "2005"
}

@article{Benincasa:2007xk,
    author = "Benincasa, Paolo and Cachazo, Freddy",
    title = "{Consistency Conditions on the S-Matrix of Massless Particles}",
    eprint = "0705.4305",
    archivePrefix = "arXiv",
    primaryClass = "hep-th",
    reportNumber = "UWO-TH-07-09",
    month = "5",
    year = "2007"
}

@article{Schuster:2008nh,
    author = "Schuster, Philip C. and Toro, Natalia",
    title = "{Constructing the Tree-Level Yang-Mills S-Matrix Using Complex Factorization}",
    eprint = "0811.3207",
    archivePrefix = "arXiv",
    primaryClass = "hep-th",
    reportNumber = "SLAC-PUB-13426, SU-ITP-08-32",
    doi = "10.1088/1126-6708/2009/06/079",
    journal = "JHEP",
    volume = "06",
    pages = "079",
    year = "2009"
}

@article{McGady:2013sga,
    author = "McGady, David A. and Rodina, Laurentiu",
    title = "{Higher-spin massless $S$-matrices in four-dimensions}",
    eprint = "1311.2938",
    archivePrefix = "arXiv",
    primaryClass = "hep-th",
    reportNumber = "PUPT-2454",
    doi = "10.1103/PhysRevD.90.084048",
    journal = "Phys. Rev. D",
    volume = "90",
    number = "8",
    pages = "084048",
    year = "2014"
}

@article{Weinberg:1965nx,
    author = "Weinberg, Steven",
    title = "{Infrared photons and gravitons}",
    doi = "10.1103/PhysRev.140.B516",
    journal = "Phys. Rev.",
    volume = "140",
    pages = "B516--B524",
    year = "1965"
}

@article{Arkani-Hamed:2017jhn,
    author = "Arkani-Hamed, Nima and Huang, Tzu-Chen and Huang, Yu-tin",
    title = "{Scattering amplitudes for all masses and spins}",
    eprint = "1709.04891",
    archivePrefix = "arXiv",
    primaryClass = "hep-th",
    reportNumber = "NCTS-TH/1714, NCTS-TH-1714",
    doi = "10.1007/JHEP11(2021)070",
    journal = "JHEP",
    volume = "11",
    pages = "070",
    year = "2021"
}

@article{Elvang:2016qvq,
    author = "Elvang, Henriette and Jones, Callum R. T. and Naculich, Stephen G.",
    title = "{Soft Photon and Graviton Theorems in Effective Field Theory}",
    eprint = "1611.07534",
    archivePrefix = "arXiv",
    primaryClass = "hep-th",
    doi = "10.1103/PhysRevLett.118.231601",
    journal = "Phys. Rev. Lett.",
    volume = "118",
    number = "23",
    pages = "231601",
    year = "2017"
}

@article{Grimus:1995zi,
    author = "Grimus, W. and Rebelo, M. N.",
    title = "{Automorphisms in gauge theories and the definition of CP and P}",
    eprint = "hep-ph/9506272",
    archivePrefix = "arXiv",
    reportNumber = "UWTHPH-1995-7",
    doi = "10.1016/S0370-1573(96)00030-0",
    journal = "Phys. Rept.",
    volume = "281",
    pages = "239--308",
    year = "1997"
}

@article{Falkowski:2020aso,
    author = "Falkowski, Adam and Machado, Camila S.",
    title = "{Soft Matters, or the Recursions with Massive Spinors}",
    eprint = "2005.08981",
    archivePrefix = "arXiv",
    primaryClass = "hep-th",
    doi = "10.1007/JHEP05(2021)238",
    journal = "JHEP",
    volume = "05",
    pages = "238",
    year = "2021"
}

@article{Alviani:2025msx,
    author = "Alviani, Edoardo and Falkowski, Adam",
    title = "{Gauge anomalies on shell and collinear factorization}",
    eprint = "2509.03368",
    archivePrefix = "arXiv",
    primaryClass = "hep-th",
    month = "9",
    year = "2025"
}

@article{Huang:2013vha,
    author = "Huang, Yu-tin and McGady, David",
    title = "{Consistency Conditions for Gauge Theory S Matrices from Requirements of Generalized Unitarity}",
    eprint = "1307.4065",
    archivePrefix = "arXiv",
    primaryClass = "hep-th",
    reportNumber = "PUPT-2450-MCTP-13-20",
    doi = "10.1103/PhysRevLett.112.241601",
    journal = "Phys. Rev. Lett.",
    volume = "112",
    number = "24",
    pages = "241601",
    year = "2014"
}

@article{Chen:2014eva,
    author = "Chen, Wei-Ming and Huang, Yu-tin and McGady, David A.",
    title = "{Anomalies without an action}",
    eprint = "1402.7062",
    archivePrefix = "arXiv",
    primaryClass = "hep-th",
    reportNumber = "PUPT-2459",
    month = "2",
    year = "2014"
}

@article{Boulanger:2000rq,
    author = "Boulanger, Nicolas and Damour, Thibault and Gualtieri, Leonardo and Henneaux, Marc",
    title = "{Inconsistency of interacting, multigraviton theories}",
    eprint = "hep-th/0007220",
    archivePrefix = "arXiv",
    reportNumber = "ULB-TH-00-14",
    doi = "10.1016/S0550-3213(00)00718-5",
    journal = "Nucl. Phys. B",
    volume = "597",
    pages = "127--171",
    year = "2001"
}

@article{Grisaru:1977kk,
    author = "Grisaru, Marcus T. and Pendleton, H. N.",
    title = "{Soft Spin 3/2 Fermions Require Gravity and Supersymmetry}",
    reportNumber = "Print-77-0154 (BRANDEIS)",
    doi = "10.1016/0370-2693(77)90383-5",
    journal = "Phys. Lett. B",
    volume = "67",
    pages = "323--326",
    year = "1977"
}

@article{Liu:2014vva,
    author = "Liu, Zheng-Wen",
    title = "{Soft theorems in maximally supersymmetric theories}",
    eprint = "1410.1616",
    archivePrefix = "arXiv",
    primaryClass = "hep-th",
    doi = "10.1140/epjc/s10052-015-3304-1",
    journal = "Eur. Phys. J. C",
    volume = "75",
    number = "3",
    pages = "105",
    year = "2015"
}

@article{Avery:2015iix,
    author = "Avery, Steven G. and Schwab, Burkhard U. W.",
    title = "{Residual Local Supersymmetry and the Soft Gravitino}",
    eprint = "1512.02657",
    archivePrefix = "arXiv",
    primaryClass = "hep-th",
    reportNumber = "BROWN-HET-1689",
    doi = "10.1103/PhysRevLett.116.171601",
    journal = "Phys. Rev. Lett.",
    volume = "116",
    number = "17",
    pages = "171601",
    year = "2016"
}

@article{Lysov:2015jrs,
    author = "Lysov, Vyacheslav",
    title = "{Asymptotic Fermionic Symmetry From Soft Gravitino Theorem}",
    eprint = "1512.03015",
    archivePrefix = "arXiv",
    primaryClass = "hep-th",
    month = "12",
    year = "2015"
}

@article{Cheung:2017pzi,
    author = "Cheung, Clifford",
    title = "{TASI Lectures on Scattering Amplitudes}",
    eprint = "1708.03872",
    archivePrefix = "arXiv",
    primaryClass = "hep-ph",
    reportNumber = "CALT-TH-2017-041",
    doi = "10.1142/9789813233348_0011",
    journal = "PoS",
    volume = "TASI2017",
    pages = "011",
    year = "2018"
}

@book{Schwartz:2013pla,
    author = "Schwartz, Matthew D.",
    title = "{Quantum Field Theory and the Standard Model}",
    isbn = "978-1-107-03473-0",
    publisher = "Cambridge University Press",
    address = "Cambridge, UK",
    year = "2014"
}

@article{Arkani-Hamed:2008bsc,
    author = "Arkani-Hamed, Nima and Kaplan, Jared",
    title = "{On Tree Amplitudes in Gauge Theory and Gravity}",
    eprint = "0801.2385",
    archivePrefix = "arXiv",
    primaryClass = "hep-th",
    doi = "10.1088/1126-6708/2008/04/076",
    journal = "JHEP",
    volume = "04",
    pages = "076",
    year = "2008"
}

@book{Eden:1966dnq,
    author = "Eden, Richard John and Landshoff, Peter V. and Olive, David I. and Polkinghorne, John Charlton",
    title = "{The analytic S-matrix}",
    isbn = "978-0-521-04869-9",
    publisher = "Cambridge Univ. Press",
    address = "Cambridge",
    year = "1966"
}

@book{Hannesdottir:2022bmo,
    author = "Hannesdottir, Holmfridur Sigridar and Mizera, Sebastian",
    title = "{What is the i{\ensuremath{\varepsilon}} for the S-matrix?}",
    eprint = "2204.02988",
    archivePrefix = "arXiv",
    primaryClass = "hep-th",
    doi = "10.1007/978-3-031-18258-7",
    isbn = "978-3-031-18257-0, 978-3-031-18258-7",
    publisher = "Springer",
    series = "SpringerBriefs in Physics",
    month = "1",
    year = "2023"
}

@article{Stapp:1962nxd,
    author = "Stapp, Henry P.",
    title = "{Derivation of the CPT Theorem and the Connection between Spin and Statistics from Postulates of the S-Matrix Theory}",
    doi = "10.1103/PhysRev.125.2139",
    journal = "Phys. Rev.",
    volume = "125",
    number = "6",
    pages = "2139",
    year = "1962"
}

@article{Weinberg:1964cn,
    author = "Weinberg, Steven",
    title = "{Feynman Rules for Any Spin}",
    doi = "10.1103/PhysRev.133.B1318",
    journal = "Phys. Rev.",
    volume = "133",
    pages = "B1318--B1332",
    year = "1964"
}

@book{Strominger:2017zoo,
    author = "Strominger, Andrew",
    title = "{Lectures on the Infrared Structure of Gravity and Gauge Theory}",
    eprint = "1703.05448",
    archivePrefix = "arXiv",
    primaryClass = "hep-th",
    isbn = "978-0-691-17973-5",
    publisher = "Princeton University Press",
    year = "2018"
}

@article{Bondi:1962px,
    author = "Bondi, H. and van der Burg, M. G. J. and Metzner, A. W. K.",
    title = "{Gravitational waves in general relativity. 7. Waves from axisymmetric isolated systems}",
    doi = "10.1098/rspa.1962.0161",
    journal = "Proc. Roy. Soc. Lond. A",
    volume = "269",
    pages = "21--52",
    year = "1962"
}

@article{Sachs:1962wk,
    author = "Sachs, R. K.",
    title = "{Gravitational waves in general relativity. 8. Waves in asymptotically flat space-times}",
    doi = "10.1098/rspa.1962.0206",
    journal = "Proc. Roy. Soc. Lond. A",
    volume = "270",
    pages = "103--126",
    year = "1962"
}

\end{document}